\documentclass[12pt]{article}

\usepackage{graphicx}
\usepackage{graphics}
\usepackage{epsfig}
\usepackage{amssymb}
\usepackage{amsmath}
\usepackage{color}
\usepackage{textcomp}
\usepackage{latexsym}

\usepackage[margin=2.5cm,footskip=0.6cm]{geometry}
\def\bea{\begin{eqnarray}}
\def\eea{\end{eqnarray}}
\def\beq{\begin{equation}}
\def\eeq{\end{equation}}

\def\bm{\begin{math}}
\def\me{\end{math}}

\def\del{\partial}
\def\grad{\nabla}

\begin{document}

\begin{center}
{\Large{\bf Surface-Directed Spinodal Decomposition on Chemically Patterned Substrates}} \\
\ \\
\ \\
by \\
Prasenjit Das$^{1,2}$, Prabhat K. Jaiswal$^3$ and Sanjay Puri$^2$ \\
$^1$Department of Chemical and Biological Physics, Weizmann Institute of Science, Rehovot 76100, Israel. \\
$^2$School of Physical Sciences, Jawaharlal Nehru University, New Delhi 110067, India. \\
$^3$Department of Physics, Indian Institute of Technology Jodhpur, Karwar 342037, India. \\
\end{center}

\begin{abstract}
\noindent{\it Surface-directed spinodal decomposition} (SDSD) is the kinetic interplay of phase separation and wetting at a surface. This process is of great scientific and technological importance. In this paper, we report results from a numerical study of SDSD on a chemically patterned substrate. We consider simple surface patterns for our simulations, but most of the results apply for arbitrary patterns. In layers near the surface, we observe a dynamical crossover from a {\it surface-registry regime} to a {\it phase-separation regime}. We study this crossover using layer-wise correlation functions and structure factors, and domain length scales.
\end{abstract}

\newpage
\section{Introduction}
\label{sec1}

Consider a homogeneous binary (AB) mixture at high temperature, which is rapidly quenched below the critical temperature $T_c$. The system becomes thermodynamically unstable, and its subsequent evolution is characterized by the emergence and growth of domains enriched in A or B. The kinetics of phase separation has been extensively studied \cite{pw09,dp04,ao02}. The \textit{Cahn-Hilliard-Cook} (CHC) model successfully describes diffusion-driven segregation in a mixture~\cite{pw09,dp04,ao02,hh77}. The equal-time {\it correlation function} [$C(\vec{r},t)$, where $\vec{r}$ is the distance] and {\it structure factor} [$S(\vec{k},t)$, where $\vec{k}$ is the wave-vector] of the order parameter field are used to study domain growth. For a translationally invariant and isotropic system, these quantities exhibit dynamical scaling as follows:
\begin{eqnarray}
\label{scale1}
C(\vec{r},t) &=& g\left(r/L\right), \\
\label{scale2}
S(\vec{k},t) &=& L^df\left(kL\right) .
\end{eqnarray}
In Eqs.~(\ref{scale1})-(\ref{scale2}), $g(x)$ and $f(p)$ are scaling functions, $L(t)$ is the time-dependent domain size, and $d$ is the dimensionality. For conserved kinetics with diffusive transport, the rate of change of $L(t)$ is proportional to the particle current $\vec{J}$, which is identified as the gradient of the chemical potential $\mu$. Thus,
\begin{eqnarray}
\label{LS}
\frac{d L}{dt} \sim J \sim \mid \vec \nabla \mu \mid \sim \frac{\sigma}{L^2} ,
\end{eqnarray}
where $\sigma$ is the A-B interfacial tension. The solution of Eq.~(\ref{LS}) yields $L(t) \sim (\sigma t)^{1/3}$, which is known as the \textit{Lifshitz-Slyozov} (LS) growth law \cite{ls61,dh86}.

In experiments, the segregating system is often confined to a container whose walls may have a preferential attraction for one of the components of the mixture (say, A). The interplay between phase-separation kinetics and wetting kinetics at a surface S is referred to as {\it surface-directed spinodal decomposition} (SDSD). This process is of great technological importance, and has attracted much experimental \cite{jnk91,gk95,gk03} and theoretical \cite{pf97,kb98,sp05,kpd10} attention. The presence of a surface or substrate breaks translational symmetry in the normal direction. As a result, different morphologies and kinetics are observed near the surface. The system shows either a {\it partially wet} (PW) or {\it completely wet} (CW) equilibrium morphology, depending upon the relative surface tensions between A, B and S. For the PW morphology, the interface between A-rich and B-rich domains makes an angle $\theta$ with the substrate S. This \textit{contact angle} is determined by Young's condition \cite{ty05}:
\begin{eqnarray}
\label{eqn1}
\sigma \cos~\theta = \gamma_{BS} - \gamma_{AS},
\end{eqnarray}
where $\gamma_{AS}$ and $\gamma_{BS}$ are surface tensions between the A-rich and B-rich phases and S, respectively. Eq.~(\ref{eqn1}) does not have a solution when $\gamma_{BS} - \gamma_{AS} > \sigma$, and the B-rich phase is completely expelled from the surface forming a CW morphology. In this case, the A-B interface is parallel to the substrate.

The problem of SDSD on chemically homogeneous and physically flat substrates has been studied extensively via experiments~\cite{jnk91,gk95,gk03} and simulations \cite{pb92,pbzp92,pb94,pbf97,pb01,be90,bc92,jm93}. The first successful coarse-grained model for SDSD was proposed by Puri and Binder (PB) \cite{pb92}, who supplemented the CHC model with two boundary conditions which modeled the surface. PB showed that, for the CW morphology, the surface gives rise to an SDSD wave. This consists of alternating wetting and depletion layers of the preferred component. The SDSD wave propagates into the system, as seen experimentally by Jones et al. \cite{jnk91}. PB focused on two experimentally relevant features of the SDSD morphology: \\
(a) the growth law for the wetting layer thickness $R_1(t)$; \\
(b) the scaling behavior of layer-wise correlation functions $C(\vec{\rho},z,t)$ and structure factors $S(\vec{k_\rho},z,t)$, where $\vec{\rho}$ and $z$ denote the coordinates parallel and perpendicular to the surface (located at $z=0$).

PB showed that $R_1(t)$ has an early-time behavior which depends on the surface potential \cite{pb01}. At late times, $R_1(t)$ shows a crossover to the universal LS behavior. They also studied the effect of off-criticality on the above picture. For moderately off-critical quenches, where the bulk exhibits spinodal decomposition (SD), the above scenario applies. However, for highly off-critical mixtures, where the homogeneous bulk remains stable, $R_1(t)$ shows a late-time diffusive behavior, $R_1(t) \sim t^{1/2}$. PB also showed that the layer-wise correlation functions and structure factors exhibit dynamical scaling. Further, the lateral domain size $L(z,t)$ follows the LS law, but the prefactor is higher near the surface. PB explained this as a consequence of the orientational effects of the layered SDSD profile at the surface.

In many applications, the substrate may be heterogeneous and patterned -- either chemically or physically. The process of SDSD on chemically patterned substrates has been very useful in, e.g., the paper industry \cite{kbw94}, lubrication \cite{jw02,rjh09}, enhanced oil recovery, tissue engineering, and bio-material development \cite{rxd04,cjt01}. One of the potential applications of such substrates is in stamping or contact printing~\cite{jdk98}, where an elastomeric stamp is used to transport the material to predefined regions. There exists a nano-transfer printing technique, which relies on tailored surface chemistries to transfer metal films from the raised areas of a stamp to a substrate when these are brought into contact~\cite{lwb02}. Further, chemically patterned substrates are used to self-assemble polymer mixtures~\cite{jzg02,kdl98,end98,bwm98}. In the pharmaceutical and cosmetics industries, microfluidic assays commonly require the formation of stable emulsions of immiscible fluids, such as oil and water~\cite{mg99,dbs98}. Further, chemically patterned substrates are useful in promoting the  mixing of immiscible fluids in microchannels \cite{kyb02,kb03}.

Given this large number of applications, it is useful to gain a good theoretical understanding of SDSD on chemically patterned substrates. This is the primary focus of the present paper. Before proceeding, it is useful to review some earlier work in this context. In conjunction with their experiments on polymer blends, Karim et al. \cite{kdl98} also reported results from a simulation of the CHC model at a surface. They showed that a modulation of the surface potential resulted in a corresponding checkerboard pattern in the segregating mixture. These authors did not study the detailed pattern dynamics and only showed typical simulation morphologies. An analogous study is due to Chen and Chakrabarti \cite{cc98}, who studied morphologies in a block copolymer (BCP) on a patterned substrate. The BCP is modeled by a simple variant of the CHC model \cite{os87,ob88}, and is characterized by mesoscale segregation, i.e., the segregating mixture freezes into an equilibrium structure with a typical length scale $L_s$. Chen-Chakrabarti studied the emergent morphologies as a function of $L_s/M$, where $M$ is the scale of the chemical pattern.

Let us also discuss some more recent numerical studies of this problem \cite{wd06,dpz13,clx13,spz16,hm17,xzw19,zl19}. Dessi et al. \cite{dpz13} and Serral et al. \cite{spz16} used cell dynamical system (CDS) models \cite{op87} to study SDSD in BCPs on patterned surfaces. (These CDS models were equivalent to the modified CHC equation studied by Chen and Chakrabarti \cite{cc98}.) These authors studied the structures emerging from the interplay of the BCP mesoscale morphology and the chemical pattern. Chen et al. \cite{clx13} used {\it self-consistent field theory} to study the self-assembly of BCPs on patterned substrates. Xiang et al. \cite{xzw19} used {\it dissipative particle dynamics} simulations to study structural transitions in BCPs on chemically patterned substrates.

The above are just a few representative studies of this problem. Most of these studies focused on classifying emergent morphologies. To the best of our knowledge, there is no detailed theoretical study of time-dependent quantities, e.g., length scales, structure factors, etc., for SDSD on chemically patterned substrates. This is surprising because most of these quantities are experimentally accessible. As a matter of fact, a quantitative analysis of the evolution is necessary for a proper understanding of this problem. This is the gap that we address here. 

In this paper, we use Langevin simulations to study SDSD on a chemically patterned substrate. In particular, we focus on the time-dependence of morphological features near the patterned substrate. This paper is organized as follows. In Sec.~\ref{sec2}, we describe our model of SDSD on a chemically patterned substrate. The detailed simulation results are presented in Sec.~\ref{sec3}. Finally, we conclude this paper with a summary and discussion in Sec.~\ref{sec4}.

\section{Modeling and Numerical Details}
\label{sec2}

We use the PB model \cite{pb92} of SDSD to study the phase separation kinetics of a binary (AB) mixture at a chemically patterned substrate. The order parameter is defined as $\psi(\vec r, t) = \rho_{\rm A}(\vec r, t) - \rho_{\rm B}(\vec r, t)$, where $\rho_{\rm A}(\vec r, t)$ and $\rho_{\rm B}(\vec r, t)$ are, respectively, the local concentrations  of A and B at position $\vec {r}$ and time $t$. The PB model consists of the CHC equation with a surface potential, which describes bulk phase separation. This is a fourth-order partial differential equation, and it must be supplemented by two boundary conditions, representing the effect of the surface.

We consider a short-ranged surface potential $V(\vec{\rho},z)$, which acts in a microscopic layer of thickness $a$ at the surface S:
\bea
V(\vec{\rho},z) &=& -h_1(\vec{\rho}), \quad z<a , \nonumber \\
&=& 0, \quad z>a .
\eea
We set $a=0$ as it is small compared to the coarse-graining scale. Then, in dimensionless units, the free-energy functional for an unstable binary mixture in contact with S is given by \cite{pf97,sp05}
\begin{eqnarray}
\label{eqn2}
\mathcal F\left[\psi (\vec r)\right] &=&{\int d\vec{\rho} \int_0^\infty dz \left[ - \frac{\psi^2}{2} + \frac{\psi^4}{4} + \frac{1}{4}(\vec{\nabla} \psi)^2 \right] } \nonumber \\
&& + {\int d\vec\rho \left[ - \frac{g}{2}\psi (\vec \rho, 0)^2 - h_1(\vec{\rho}) \psi (\vec \rho,0) - \gamma\psi (\vec \rho,0)\frac{\del \psi}{\del z}\Big|_{z=0} + \frac{\tilde{\gamma}}{2} \left(\vec{\nabla}_\rho
\psi (\vec \rho, 0) \right)^2 \right]} \nonumber \\
& \equiv & F_b + F_s.
\end{eqnarray}
Here, we have decomposed coordinates as $\vec r = (\vec\rho, z)$, as mentioned earlier. The dimensionless rescaling is provided in Ref.~\cite{sp05}. Nevertheless, for ease of reference, it is useful to recall some details here. The order-parameter scale is
\beq
\psi_0 = \sqrt{3} \left(\frac{T_c}{T} -1\right)^{1/2} ,
\eeq
where $T_c$ and $T$ are the critical and quench temperatures, respectively. The length scale is the bulk correlation length:
\beq
\xi_b = \left[\frac{q}{2} \left(1- \frac{T}{T_c}\right)\right]^{-1/2} ,
\eeq
where $q$ is the {\it coordination number} of the system.

\begin{figure}[t]
\centering
\includegraphics*[width=0.4\textwidth]{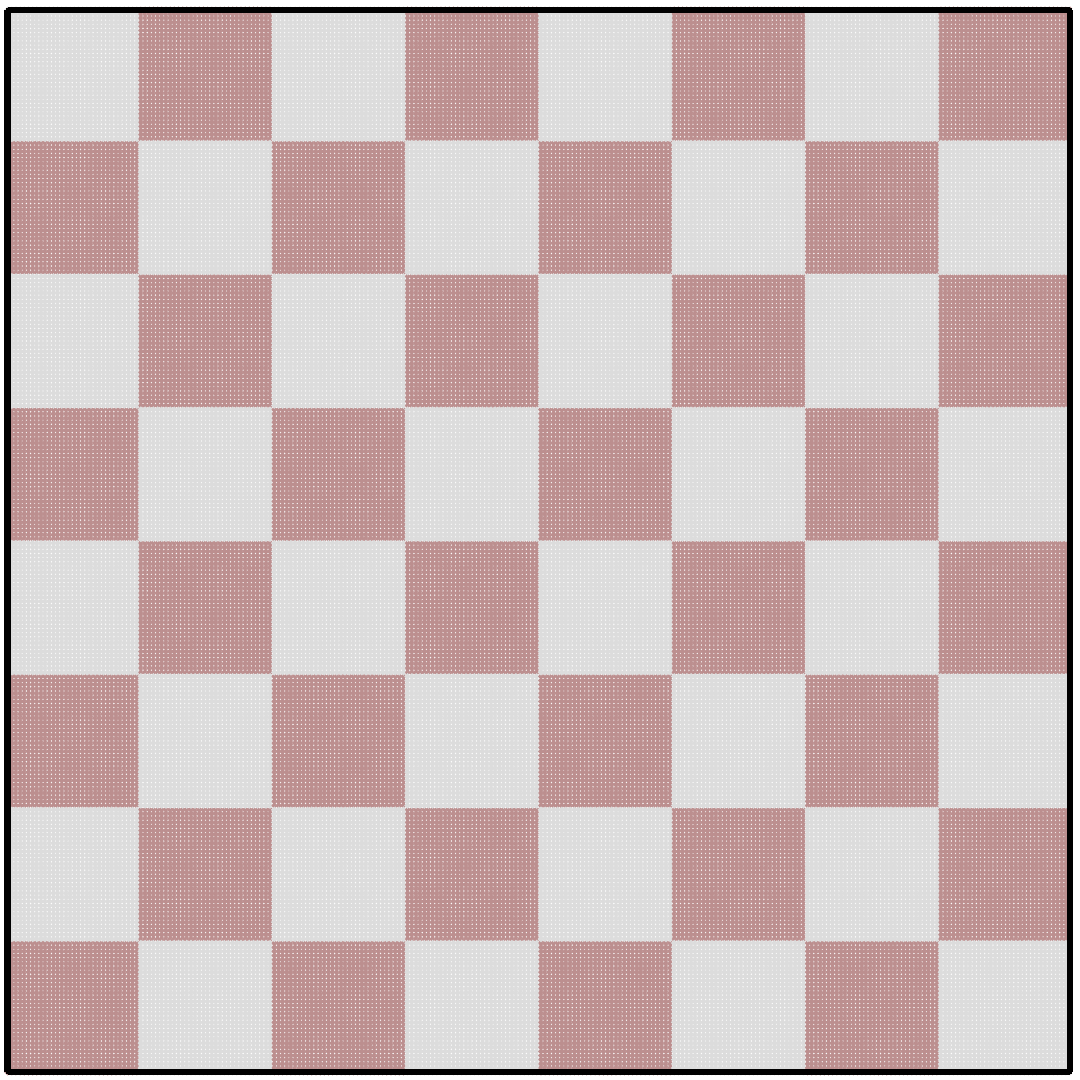}
\caption{\label{f1} (color online) Schematic of a chemically patterned substrate. The substrate is decorated with a checkerboard pattern of brown and gray patches, which are chemically distinct. Each brown (gray) patch of the substrate preferably attracts the A (B) component of a binary (AB) mixture.}
\end{figure}

In Eq.~(\ref{eqn2}), $F_b$ has the usual $\psi^4$-form of the bulk free energy \cite{pw09,dp04}. The term $F_s$ is the contribution to the free energy from the surface. In $F_s$, the phenomenological constants $g, \gamma, \tilde{\gamma}$ are related to the bulk correlation length and other system parameters \cite{pf97,sp05}. For simplicity, we assume that the free-energy cost of the $(\vec{\grad} \psi)^2$-term is the same in the bulk and surface layers, i.e., we set $\tilde{\gamma} = 0.5$. In real experiments, these may differ somewhat but would have the same order of magnitude. The one-sided derivative term $\del \psi/\del z |_{z=0}$ appears due to the absence of neighboring atoms for $z<0$. The chemical pattern on the surface is reflected in the $\vec{\rho}$-dependence of $h_1(\vec{\rho})$. If $h_1(\vec{\rho}) > 0$, the surface attracts A, and $h_1(\vec{\rho}) < 0$ means that the surface is wetted by B. Figure~\ref{f1} is a schematic of a checkerboard substrate, where chemically
distinct patches are marked in different colors.

The evolution of $\psi(\vec r, t)$ in the bulk is described by the CHC equation~\cite{pw09,dp04} as follows:
\begin{eqnarray}
\label{eqn3}
\frac{\del\psi(\vec r, t)}{\del t} = -\vec{\nabla} \cdot \vec{J}(\vec{r},t) = \vec\nabla\cdot\left[\vec\nabla\left(\frac{\delta \mathcal F}{\delta \psi}\right) +
\vec \theta(\vec r, t)\right] ,
\end{eqnarray}
where $\vec\theta(\vec r, t)$ is a vector Gaussian white noise. The noise has zero average, and obeys the fluctuation-dissipation relation:
\begin{eqnarray}
\label{eqn4}
\overline{\theta_i(\vec r,t)} &=& 0~~~\forall~~~i , \\
\label{eqn5}
\overline{\theta_i(\vec r^{\,\prime}, t^\prime)\theta_j(\vec r^{\,\prime\prime}, t^{\prime\prime})} &=& 2 \epsilon \delta_{ij} \delta(\vec r^{\,\prime} - \vec r^{\,\prime\prime})\delta(t^\prime - t^{\prime\prime}) .
\end{eqnarray}
Note that this is the usual conserved noise as it has been added to the current term \cite{hh77}. An equivalent formulation is to define the noise as $\eta = \vec{\nabla} \cdot \vec{\theta}$. In that case, Eq.~(\ref{eqn5}) would have an extra Laplacian operator on the right-hand-side. Here, $\epsilon$ characterizes the strength of the noise in dimensionless units. It is related to the temperature as \cite{sp05}
\beq
\epsilon = \frac{1}{3} \left(\frac{T_c}{T} -1\right)^{-2} \xi_b^{-d} .
\eeq
For $d < 4$, $\epsilon$ diverges as $T \rightarrow T_c^-$ so that order is destroyed at the critical temperature. For bulk phase separation, the asymptotic pattern dynamics is not affected by the noise amplitude. This is because thermal fluctuations only increase the thickness of the interfaces, which we denote as $w$. However, the asymptotic regime is realized when $w/L(t) \rightarrow 0$. Thus, the presence of noise only delays the onset of the asymptotic scaling behavior \cite{po88}.

Using Eqs.~(\ref{eqn2}) and (\ref{eqn3}), we obtain 
\begin{eqnarray}
\label{eqn6}
\frac{\del\psi(\vec r, t)}{\del t} = \vec\nabla\cdot\left[\vec\nabla\left\{- \psi + \psi^3 - \frac{1}{2}\nabla^2 \psi \right\} + \vec\theta (\vec r, t)\right] , \quad z>0 .
\end{eqnarray}
The corresponding (dimensionless) boundary conditions proposed by PB at the surface are as follows:
\begin{eqnarray}
\label{eqn7}
\tau_0 \frac{\del\psi(\vec\rho, 0, t)}{\del t} &=& -\frac{\delta \mathcal F}{\delta \psi(\vec \rho , 0,t)} \nonumber \\
&=& h_1(\vec{\rho}) + g\psi(\vec \rho , 0,t) + \gamma \frac{\del \psi}{\del z}\Big |_{z=0} + \tilde{\gamma}\nabla^2_\rho \psi(\vec \rho, 0, t) , \\
\label{eqn8}
0 &=& \left[\frac{\del}{\del z}\left\{ - \psi + \psi^3 - \frac{1}{2}\nabla^2\psi \right\} +
\theta_z\right]_{z=0} ,
\end{eqnarray}
where $\tau_0$ is a relaxational time-scale. Equation~(\ref{eqn7}) describes nonconserved relaxational kinetics of the order parameter at the substrate, and rapidly relaxes the order parameter to its surface value. It can also be replaced by its static counterpart with $\del \psi/\del t = 0$. Equation~(\ref{eqn8}) sets the $z$-component of current at the surface to zero, as there is no flux across the substrate. The quantities $h_1(\vec{\rho}), g,\gamma$ and $\tilde{\gamma}$ determine the equilibrium phase diagram of the system~\cite{pf97,sp05}.

In Sec.~\ref{sec3}, we will present results for SDSD in a critical AB mixture on a chemically patterned substrate in $d=2,3$. In $d=2$, the linear substrate is placed at $z=0$, and consists of alternating chemically distinct patches of size $M_x$ which are wetted by A and B, respectively. Similarly, in $d=3$, a checkerboard substrate (cf. Fig.~\ref{f1}) is placed in the $(x,y)$-plane at $z=0$. It consists of alternating rectangular patches of size $M_x\times M_y$. For the $d=3$ case, we will also briefly discuss SDSD on a random substrate. In this case, the substrate gives rise to a random field $h_1(\vec{\rho})$ due to the presence of surface impurities. Such substrates are common in natural systems.

Using the Euler-discretization technique, we numerically solved Eqs.~(\ref{eqn6})-(\ref{eqn8}). The discretization mesh sizes were $\Delta x = 1.0$ and $\Delta t = 0.03$, which give a stable numerical scheme. The lattice size was $L_x \times L_z$ in $d=2$ ($L_x = 1024, L_z = 256$), and $L_x \times L_y \times L_z$ in $d=3$ ($L_x = L_y = 256, L_z = 64$). The boundary conditions in Eqs.~(\ref{eqn7})-(\ref{eqn8}) were imposed at $z=0$. We used free boundary conditions at $z=L_z$:
\begin{eqnarray}
&& 0 = \frac{\del\psi}{\del z} \bigg|_{z=L_z} , \\
&& 0 = \left[\frac{\del}{\del z}\left\{ - \psi + \psi^3 - \frac{1}{2}\nabla^2\psi \right\} + \theta_z\right]_{z=L_z} .
\end{eqnarray}
We imposed periodic boundary conditions in all other directions.

We started the simulation with a random initial condition for the order parameter: $\psi(\vec r, 0)= 0.0 \pm 0.01$. This mimics the disordered state with critical composition, prior to the quench below $T_c$. The noise amplitude $\epsilon$ = 0.041, corresponding to deep quenches with $T \simeq 0.22T_c$ \cite{pf97,sp05}. The lateral diffusion coefficient in Eq.~(\ref{eqn7}) is $\tilde{\gamma}=0.5$. The other parameter values were chosen as follows: \\
(a) For the checkerboard substrate, the chemically distinct patches attracted A (B) with $h_1=+1.0$ ($-1.0$). Further, $g=-0.4$ and $\gamma=+0.4$. This corresponds to the CW morphology in equilibrium for a homogeneous substrate. \\
(b) For the random substrate, the field $h_1(\vec{\rho})$ was a random variable chosen from a Gaussian distribution with average 0 and variance $\Delta = 2.0$. Moreover, $g=-0.4$ and $\gamma=+0.4$.

\section{Detailed Numerical Results}
\label{sec3}

\subsection{Checkerboard Substrates}

\begin{figure}[t]
\centering
\includegraphics*[width=0.50\textwidth]{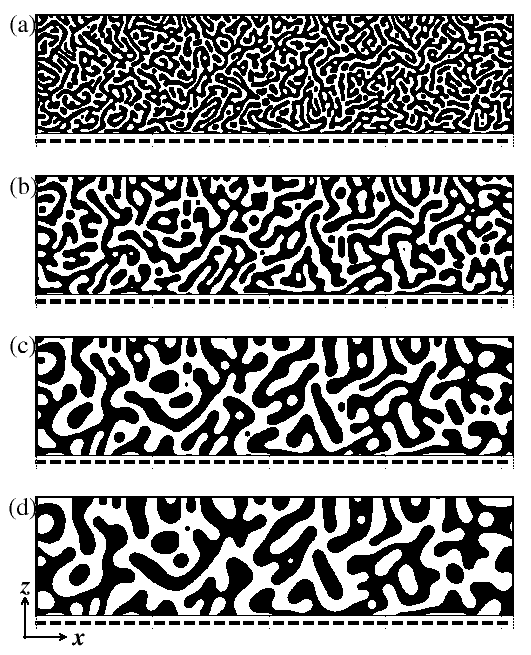}
\caption{\label{f2} Snapshots of SDSD in an unstable binary (AB) mixture evolving from a homogeneous initial condition with critical composition ($\psi_0$ = 0). We show snapshots at (a) $t=300$, (b) $t=1800$, (c) $t=5400$, (d) $t=10800$. The A-rich regions with $\psi > 0$ are marked in black, while the B-rich regions with $\psi < 0$ are unmarked. The snapshots correspond to a $d=2$ lattice of size $L_x \times L_z=1024\times 256$. The length of each chemical patch is $M_x=16$ -- the regions which attract A are marked by lines just below $z=0$.}
\end{figure}

In this paper, we will primarily focus on checkerboard substrates. Let us first present representative results for SDSD in $d=2$. Figure~\ref{f2} shows the evolution snapshots of the system at different times for patches with
$M_x=16$. The linear stability analysis (about $\psi^* = 0$) of Eq.~(\ref{eqn6}) with $\vec{\theta}=0$ shows us that the most unstable wave-vector for the CHC equation is $k_m=1$, with wavelength $\lambda_m = 2\pi/k_m = 2 \pi$. (Recall that all lengths are measured in units of $\xi_b$, the bulk correlation length.) Thus, the patches in Fig.~\ref{f2} are $16/(2\pi) \simeq 2.55$ times the bulk spinodal wavelength. The domain morphology in the vicinity of the substrate is complicated. The surface pattern is always maintained in the $z=1$ layer, and the bulk is characterized by the usual bicontinuous SD morphology. In the early stages of evolution, a few layers close to the substrate maintain surface registry, but this dissolves at later times when the bulk length scale $L(t) \geq M_x$. Thus, the time-scale on which the registry melts is $t_c \sim M_x^3$. Moreover, the degree of surface registry is lower for larger $z$, due to the interference of oppositely-oriented SDSD waves originating from the checkerboard surface.

This evolution should be contrasted with SDSD at a chemically homogeneous substrate. In that case, at early times, the SDSD waves have an oscillatory profile propagating into the bulk. At later times, bulk phase separation destroys the oscillatory profile, and only the wetting layer and depletion layer survive \cite{pb92,pbf97,pb01}.

\begin{figure}[t]
\centering
\includegraphics*[width=0.55\textwidth]{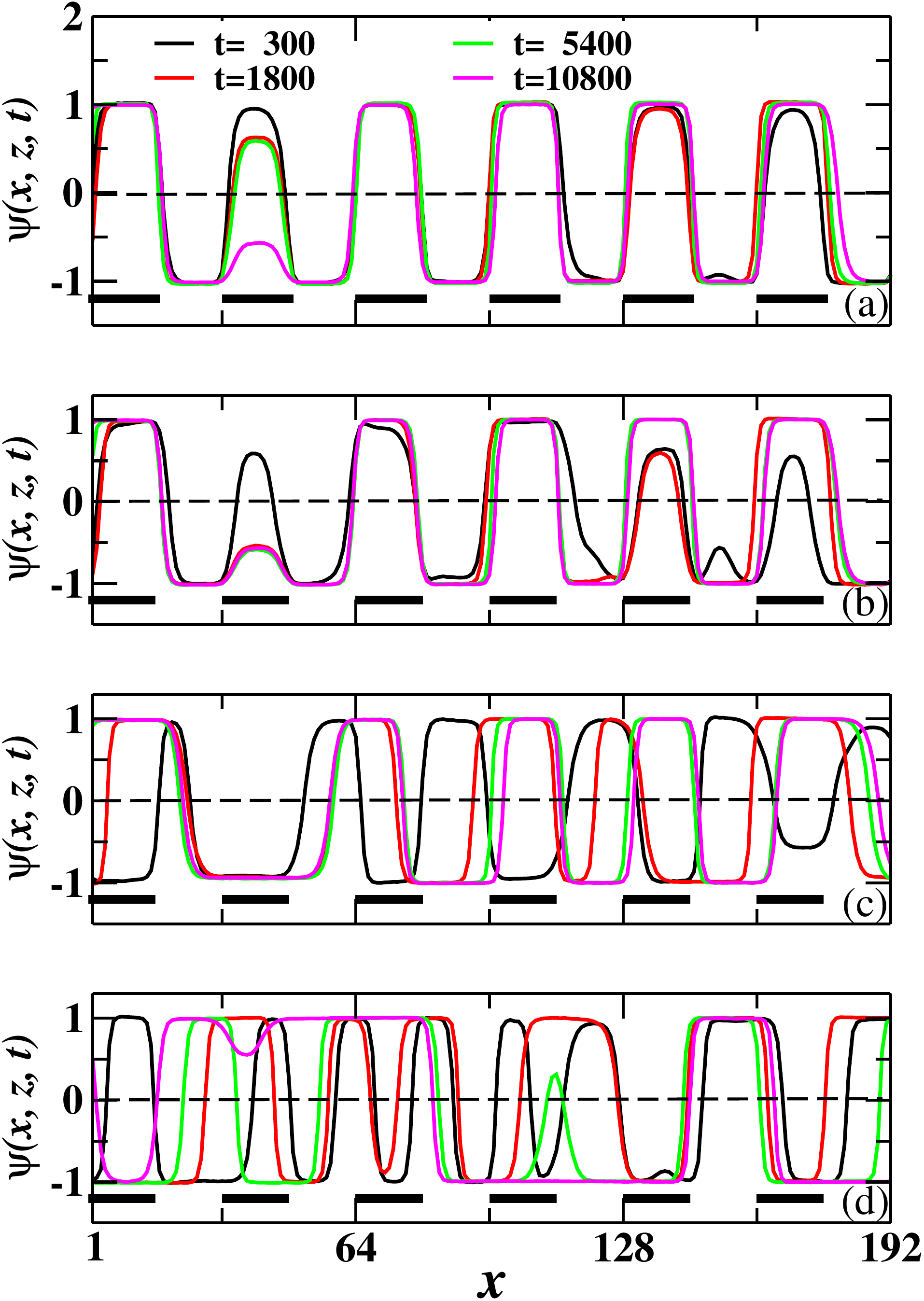}
\caption{\label{f3} (color online) Order-parameter profiles $\psi (x,z,t)$ vs. $x$ for different $t$, as specified. We show data for (a) $z=2$, (b) $z=3$, (c) $z=7$, (d) $z=128$. These profiles are obtained from the snapshots shown in Fig.~\ref{f2}. For the sake of clarity, we have shown only the region $x \in [1,192]$.}
\end{figure}

In Fig.~\ref{f3}, we plot order-parameter profiles $\psi (x,z,t)$ vs. $x$ for different $z$. These are obtained
from the snapshots in Fig.~\ref{f2}. First, consider the profile at $z=2$. For $t \leq 2000$, the profile exhibits an alternating behavior imposed by the surface pattern. This melts at later times, as the bulk segregation becomes the dominant process. A similar statement applies for $z=3$, except that we already see the initial stages of melting of the registry by $t=300$ (the earliest time shown). The profiles at $z=128$ show no signs of the surface pattern, as expected. They are just the usual bulk SD profiles.

\begin{figure}[t]
\centering
\includegraphics*[width=0.50\textwidth]{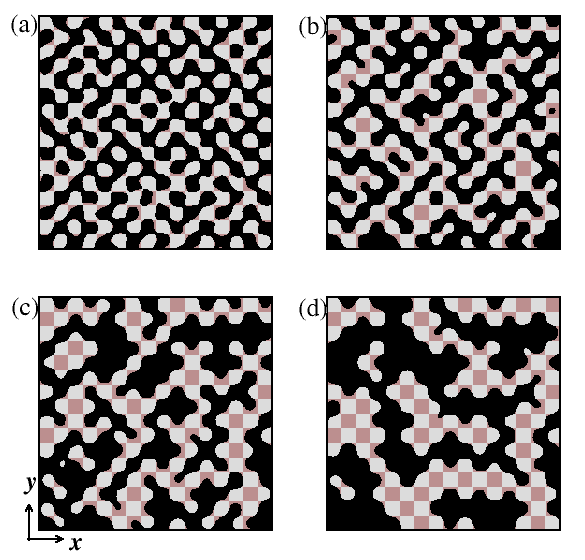}
\caption{\label{f4} (color online) Snapshots of SDSD in an unstable binary (AB) mixture evolving from a homogeneous initial condition with critical composition ($\psi_0$ = 0) in $d=3$. The snapshots correspond to (a) $t=270$, (b) $t=2700$, (c) $t=5400$, (d) $t=10800$. The system size is $L_x\times L_y\times L_z=256\times256\times64$. The snapshots show a cross-section in the $(x,y)$-plane at $z=3$. The projections of chemically distinct patches on the substrate are marked in brown and gray, respectively. The size of the patches is $M_x\times M_y= 16 \times 16$.}
\end{figure}

We next present results for the $d=3$ case, where the surface is 2-dimensional. In Fig.~\ref{f4}, we show
evolution snapshots of the $z=3$ layer. The size of patches on the substrate is $M_x\times M_y=16 \times 16$. (As in the $d=2$ case, the $z=1$ layer is in registry with the substrate.) We observe a checkerboard morphology at $t=270$. However, the domains are circular and form connecting necks to minimize the surface tension. At later times, this morphology starts melting as bulk phase separation dominates over surface-field-driven patterning. The snapshot at $t=10800$ shows that the checkerboard morphology has almost completely disappeared. The only remaining sign of the surface pattern is the corrugated structure on the domain boundaries. (In bulk SD, the interfaces are smooth and flat as this minimizes curvature.) As in the $d=2$ case, the persistence time of the registry scales as $t_c \sim M_x^3$. Clearly, $t_c$ also increases with $h_1$, the strength of the surface field.

\begin{figure}[t]
\centering
\includegraphics*[width=0.50\textwidth]{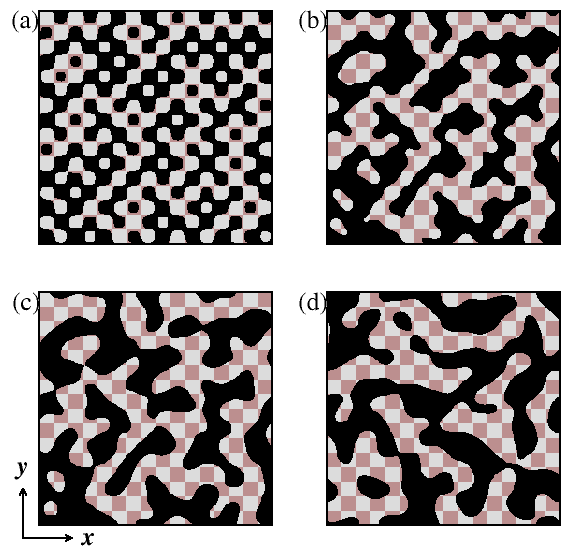}
\caption{\label{f5} (color online) Snapshots at $t=5400$ for (a) $z=2$, (b) $z=5$, (c) $z=10$, (d) $z=32$. The other details are the same as in Fig.~\ref{f4}.}
\end{figure}

In Fig.~\ref{f5}, to study the role of $z$, we plot different layers at $t=5400$. As we move away from the
substrate, the checkerboard morphology disappears. The domain morphology at $z=10$ barely shows any sign of the surface pattern. For a given patch size, the depth of surface registry increases with $h_1$.

\begin{figure}[t]
\centering
\includegraphics*[width=0.70\textwidth]{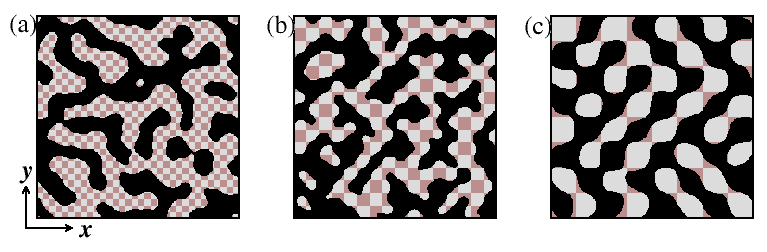}
\caption{\label{f6} (color online) Snapshots at $t=5400$ and $z=5$ for patch sizes (a) $M_x\times M_y=8\times 8$, (b) $M_x\times M_y=16\times 16$,  (c) $M_x\times M_y=32\times 32$.}
\end{figure}

Further, to study the effect of patch size on SDSD, we plot cross-sections in the $(x,y)$-plane at a given height, $z=5$. Figure~\ref{f6} shows evolution snapshots at $t=5400$ for different patch sizes. Recall that the surface
registry is destroyed when the bulk length scale $L(t)$ becomes comparable to the patch size. Therefore, at a given height, we expect the registry to melt earlier for smaller $M_x$. For patch sizes $8^2$, the morphology is similar to that for bulk phase separation, and independent of the surface pattern. However, for patch sizes $16^2$, the pattern is intermediate between registry and bulk SD. The domain boundaries are affected by the patterning on the substrate. Finally, the checkerboard morphology persists at $t=5400$ for patch size $32^2$.

The primary aim of this paper is to make quantitative statements about the time-dependence of the domain morphology. Let us now tackle this task. In order to characterize the morphologies \cite{pb92}, we calculated the layer-wise correlation function $C(\vec{\rho},z,t)$ of the order parameter field. Here, we choose layers close to the substrate, as well as in the bulk of the system. The layer-wise equal-time correlation function is defined as
\begin{eqnarray}
\label{eqn9}
C(\vec{\rho}, z, t) = \frac{1}{L_x\times L_y} \int d\vec R\left[\langle \psi(\vec R, z, t) \psi(\vec R+\vec\rho, z, t)\rangle - \langle\psi(\vec R, z, t)\rangle \langle\psi(\vec R+\vec\rho, z, t)\rangle\right] ,
\end{eqnarray}
where the angular brackets denote an averaging over independent initial conditions and thermal fluctuations. The pattern is isotropic in the $\vec{\rho}$-plane, so we spherically average $C(\vec{\rho},z,t)$ to obtain
$C(\rho,z,t)$. If the evolution of the system is characterized by a single $z$-dependent length scale $L(z,t)$, we expect the correlation functions to exhibit dynamical scaling:
\beq
\label{cads}
C(\rho,z,t) = g_z \left[\frac{\rho}{L(z,t)}\right] .
\eeq
Eq.~(\ref{cads}) is the generalization of Eq.~(\ref{scale1}), where we allow for the possibility that the scaling function also depends on $z$.

We also compute the layer-wise structure factor $S(\vec{k}_\rho,z,t)$, which is the Fourier transform of $C(\vec{\rho},z,t)$ at wave vector $\vec k_\rho$:
\begin{eqnarray}
\label{eqn10}
S(\vec{k}_\rho,z,t) = \int d\vec\rho~e^{i\vec{k_\rho} \cdot \vec\rho}~C\left(\vec\rho, z, t\right) .
\end{eqnarray}
We spherically average $S(\vec{k}_\rho,z,t)$ in the $\vec{k}_\rho$-plane to obtain $S(k_\rho,z,t)$. The dynamical scaling form of $S(k_\rho,z,t)$ is the appropriate generalization of Eq.~(\ref{scale2}). 
\beq
S(k_\rho,z,t) = L(z,t)^d f_z \left[k_\rho L(z,t) \right] .
\eeq
All statistical quantities presented here were obtained as averages over 20 independent runs. Each run started from a different initial condition and had a different noise realization. As this is a nonequilibrium system, we do not average statistical quantities over time.

\begin{figure}[t]
\centering
\includegraphics*[width=0.35\textwidth]{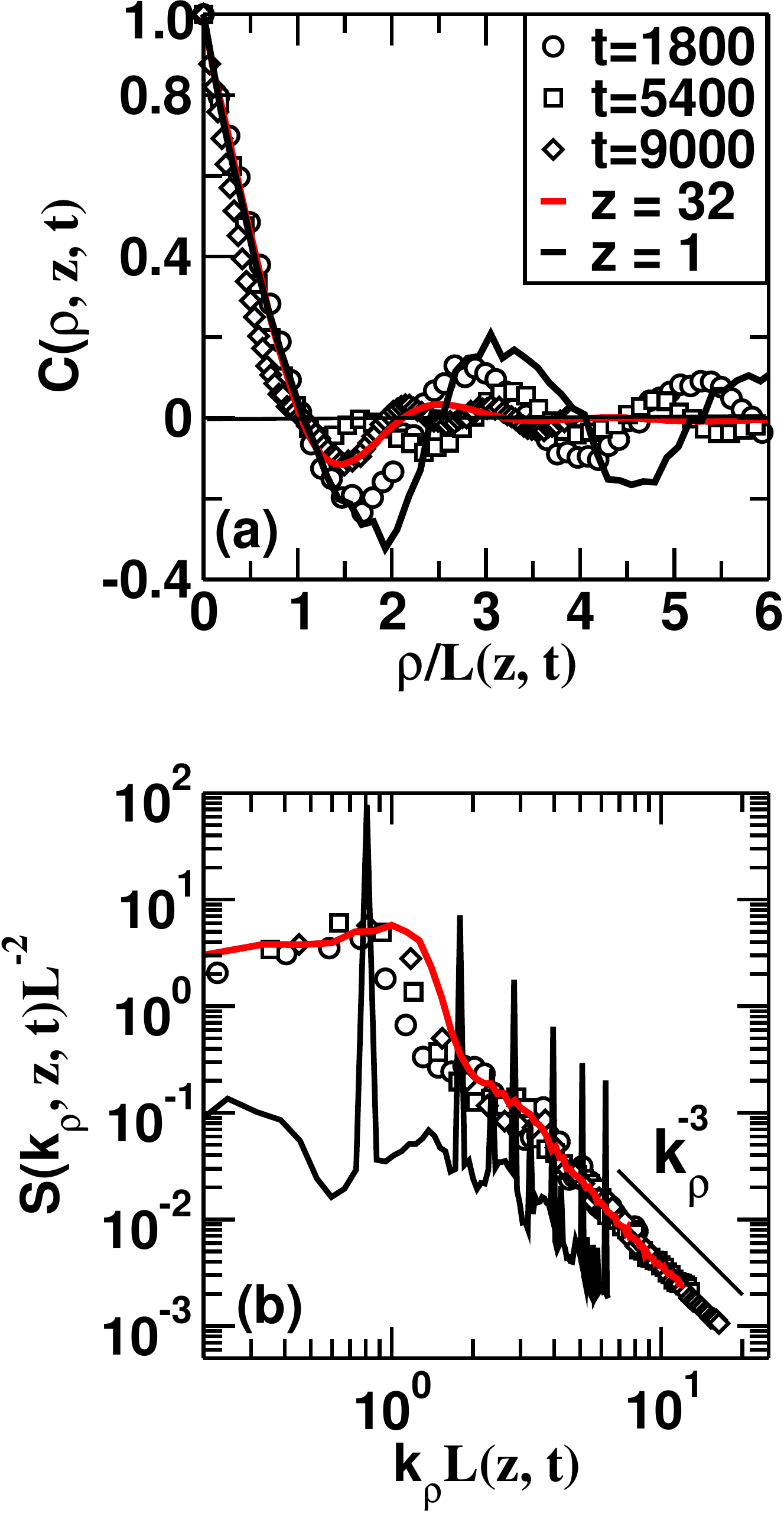}
\caption{\label{f7} (color online) Scaling plot of layer-wise correlation functions and structure factors for the evolution shown in Fig.~\ref{f4}. (a) Plot of $C(\rho,z,t)$ vs. $\rho/L(z,t)$ for $z=3$ at different times. For reference, we also show data for $z=1,32$. We define the length scale $L(z,t)$ as the first zero crossing of $C(\rho,z,t)$. (b) Log-log plot of $S(k_\rho,z,t)L(z,t)^{-2}$ vs. $k_\rho L(z, t)$ for $z=3$ at different times. We also show data for $z=1,32$. The symbols used have the same meaning as in (a). The solid line labeled $k_\rho^{-3}$ shows \textit{Porod's law}.}
\end{figure}

Fig.~\ref{f7} is a scaling plot of 
$C(\rho,z,t)$ and $S(k_\rho,z,t)$ at $z=3$ (cf. Fig.~\ref{f4}), i.e., a layer close to the substrate. In Fig.~\ref{f7}(a), we plot $C(\rho,z,t)$ vs. $\rho/L(z,t)$ at different times. We define $L(z,t)$ as the first zero crossing of $C(\rho,z,t)$. We do not observe a data collapse for $C(\rho,z,t)$, showing the breakdown of dynamical scaling in this case. This is due to the kinetic interplay of surface-field-driven registry and bulk phase separation near the substrate. This interplay results in a crossover from a checkerboard morphology (for $t < t_c$) to an SD morphology (for $t > t_c$). Therefore, $C(\rho,z,t)$ at early times [$t=1800$ in Fig.~\ref{f7}(a)] is similar to that for $z=1$ [solid line in Fig.~\ref{f7}(a)], which is a long-range oscillatory function due to the checkerboard morphology. In the bulk (for large $z=32$), we found that $C(\rho,z,t)$ obeys dynamical scaling, as the surface is not relevant. The bulk $C(\rho,z,t)$ is also shown as a solid red line in Fig.~\ref{f7}(a). We see that $C(\rho, z, t)$ at later times [$t=9000$ in Fig.~\ref{f7}(a)] is similar to that for the bulk. In Fig.~\ref{f7}(b), we plot $S(k_\rho,z,t)L^{-2}$ vs. $k_\rho L(z,t)$ on a log-log scale. As in Fig.~\ref{f7}(a), the data does not collapse but rather shows a crossover from surface registry to bulk phase separation. In the limit of large $k_\rho$, $S(k_\rho,z,t) \sim k_\rho^{-3}$ at all times, following the well-known \textit{Porod's law} \cite{gp82,op88}. This is a consequence of scattering from sharp interfaces, which are always present in the morphology even though it is undergoing a crossover in this time-window. The behavior seen in Fig.~\ref{f7} applies for all values of $z$ near the substrate, with an appropriate shift in the crossover time $t_c$.

\begin{figure}[t]
\centering
\includegraphics*[width=0.35\textwidth]{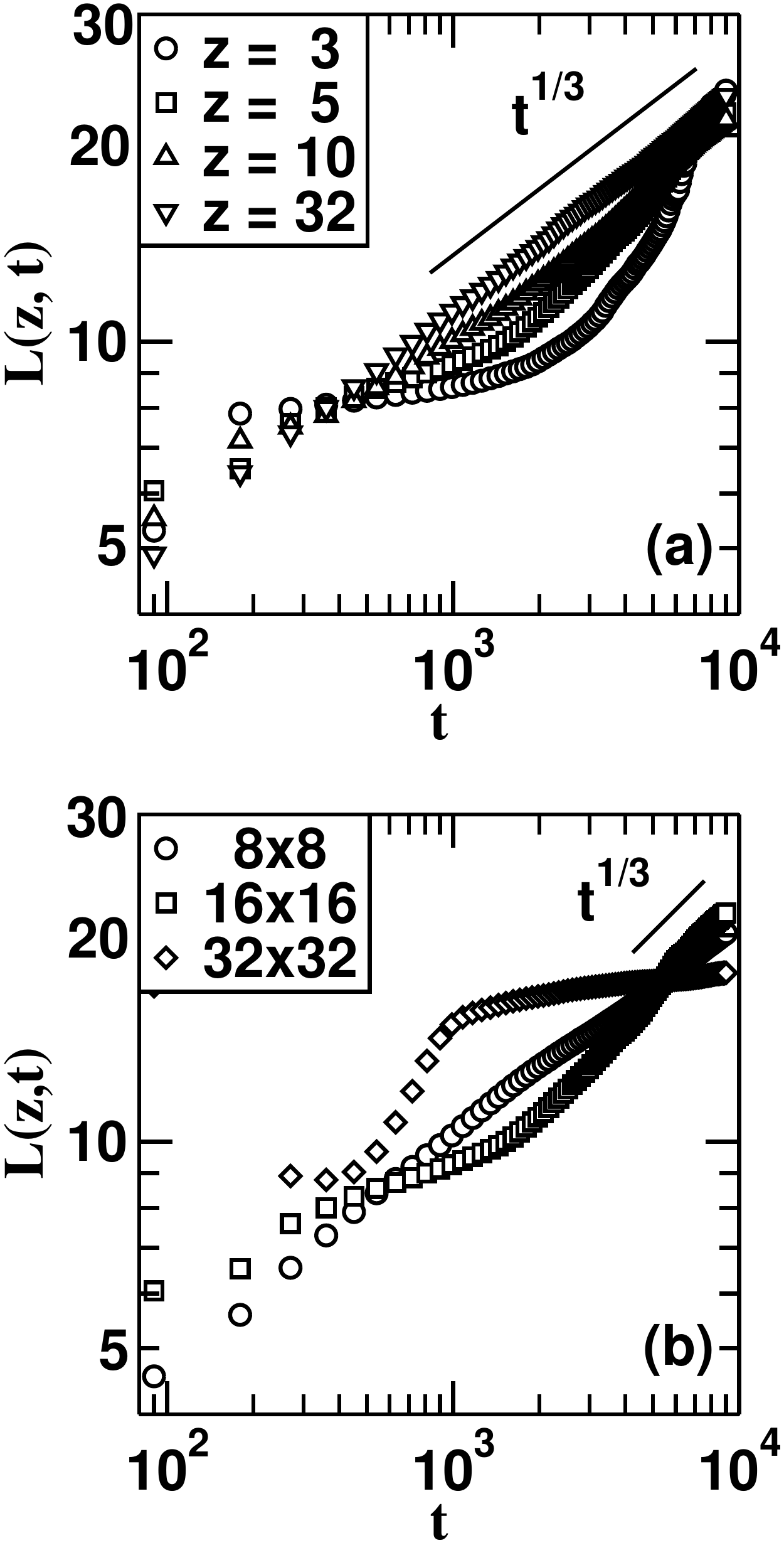}
\caption{\label{f8} Time-dependence of the characteristic length scale $L(z,t)$. (a) Log-log plot of $L(z, t)$ vs. $t$ for different $z$, as indicated. The size of patches on the substrate is $M_x\times M_y= 16\times 16$ (see Fig.~\ref{f4}). (b) Log-log plot of $L(z,t)$ vs. $t$ for $z=5$ and different patch sizes, as specified (see Fig.~\ref{f6}). The line labeled $t^{1/3}$ denotes the LS law.}
\end{figure}

Finally, we study the domain growth laws. In Fig.~\ref{f8}(a), we plot $L(z, t)$ vs. $t$ on a log-log scale for 
different values of $z$ and patch size $16^2$ (cf. Fig.~\ref{f4}). For layers far from the substrate (e.g., $z=32$), $L(z,t)$ follows the LS growth law, $L(t) \sim t^{1/3}$. We observe an anomalous growth law at early times for layers closer to the substrate, e.g., $z=3,5,10$. For an extended period of time, $L(z,t)$ is approximately constant at the patch length scale $M_x$. With the passage of time, the checkerboard is destroyed and the length scale crosses over to the LS growth law. The melting of the registry occurs layer-wise, so the data for $z=3$ is the last data set to cross over into the LS regime. To study the role of patch sizes, we plot $L(z,t)$ vs. $t$ at $z=5$ for different patch sizes in Fig.~\ref{f8}(b) (cf. Fig.~\ref{f6}). Clearly, anomalous growth is observed at early times, which crosses over to LS growth at later times. The crossover time $t_c$ increases with the patch size. As a matter of fact, our data for $M_x \times M_y = 32^2$ in Fig.~\ref{f8}(b) has not yet entered the LS regime.

In the above discussion, we have restricted ourselves to surfaces with square patches. There are also many applications which involve rectangular patches with $M_x \neq M_y$. In particular, the case of a 1-dimensional surface pattern (stripes) with $M_y = L_y$ is experimentally very interesting. We defer a study of this problem to future work.

\subsection{Random Substrates}

Next, let us present some results for randomly-patterned substrates. In many natural systems, surface impurities give rise to random fields. We model this by assuming that $h_1(\vec{\rho})$ is a random variable drawn from a Gaussian distribution:
\begin{equation}
P(h_1) = \frac{1}{\sqrt{2\pi}\Delta}e^{-h_1^2/(2\Delta^2)} ,
\end{equation}
where $\Delta$ measures the disorder strength. The results presented below correspond to $\Delta = 2.0$.

\begin{figure}[t]
\centering
\includegraphics*[width=0.80\textwidth]{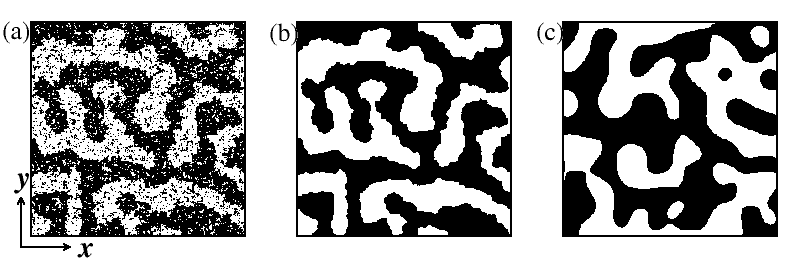}
\caption{\label{f9} Snapshots at $t=5400$ for SDSD on a random substrate. The cross-sections are taken in the $(x,y)$-plane at (a) $z=1$, (b) $z=2$, (c) $z=32$.}
\end{figure}

In Fig.~\ref{f9}, we show snapshots of the $\psi$-field in the $(x,y)$-plane for SDSD with a random field. The  
pictures correspond to $t=5400$ and $z=1,2,32$. The bulk snapshot for $z=32$ is shown for reference purposes. The snapshots for $z=1$ (surface layer) and $z=2$ show a similar morphology, as expected. However, there are two important differences: \\
(a) The domains for $z=1$ are much noisier, with many A atoms lying inside the B-rich domains and vice versa, due to the random field. \\
(b) The domain boundaries for $z=1$ are much rougher, as the interfaces find locally favorable positions in the random field. \\
These ``fractal'' domains for $z=1$ have important implications for the correlation function and structure factor, as we will see next.

\begin{figure}[t]
\centering
\includegraphics*[width=0.45\textwidth]{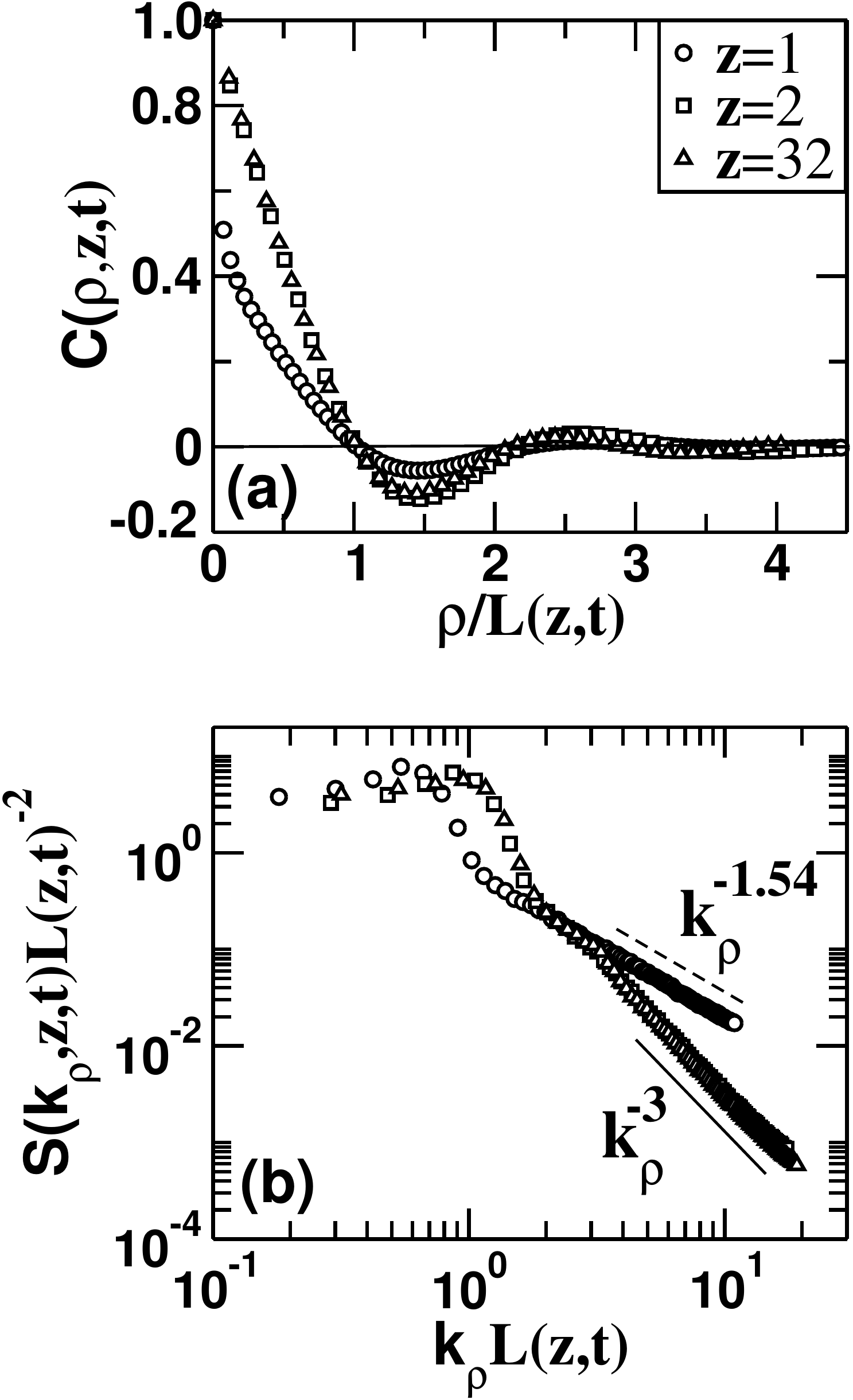}
\caption{\label{f10}Scaling plot of layer-wise correlation functions and structure factors for the snapshots shown in Fig.~\ref{f9}. (a) Plot of $C(\rho,z,t)$ vs. $\rho/L(z,t)$ for $z=1,2,32$. (b) Log-log plot of $S(k_\rho,z,t)L(z,t)^{-2}$ vs. $k_\rho L(z, t)$ for $z=1,2,32$. The dashed line labeled $k_\rho^{-1.54}$ shows a \textit{fractal Porod law}.}
\end{figure}

In Fig.~\ref{f10}, we show the scaling behavior of $C(\rho,z,t)$ and $S(k_\rho,z,t)$. In Fig.~\ref{f10}(a), the 
scaled correlation function is comparable for $z=2$ and $z=32$, showing that the bulk morphology sets in by $z=2$. However, $C(\rho,z,t)$ for $z=1$ is markedly different. An important difference is that $C(\rho,z,t)$ decays from its maximum value with a cusp behavior [$1-C(\rho,z,t) = a \rho^\theta + \ldots$]. This should be contrasted with the linear decay [$1-C(\rho,z,t) = b \rho + \ldots$] for $z=2,32$. The linear decay characterizes sharp interfaces and gives rise to a Porod law \cite{gp82} in the structure factor, $S(k_\rho,z,t) \sim k_\rho^{-(d+1)}$, which we have mentioned earlier. On the other hand, the cusp behavior for $z=1$ gives rise to a {\it fractal Porod law} \cite{bs84,db00,skb11,skb14}: $S(k_\rho,z,t) \sim k_\rho^{-(d+\theta)}$, where $\theta$ is related to the fractal dimension $d_m$ as $\theta = d_m-d$ for a mass fractal. This is confirmed in Fig.~\ref{f10}(b), where we plot $S(k_\rho,z,t) L(z,t)^{-2}$ vs. $k_\rho L(z,t)$. We see a Porod tail in the data for $z=2,32$. On the other hand, we see a fractal Porod tail in the data for $z=1$. The corresponding decay exponent is estimated as $d+\theta \simeq 1.54$, showing that the domains in Fig.~\ref{f9}(a) are mass fractals with $d_m \simeq 1.54$. 

\begin{figure}
\centering
\includegraphics*[width=0.50\textwidth]{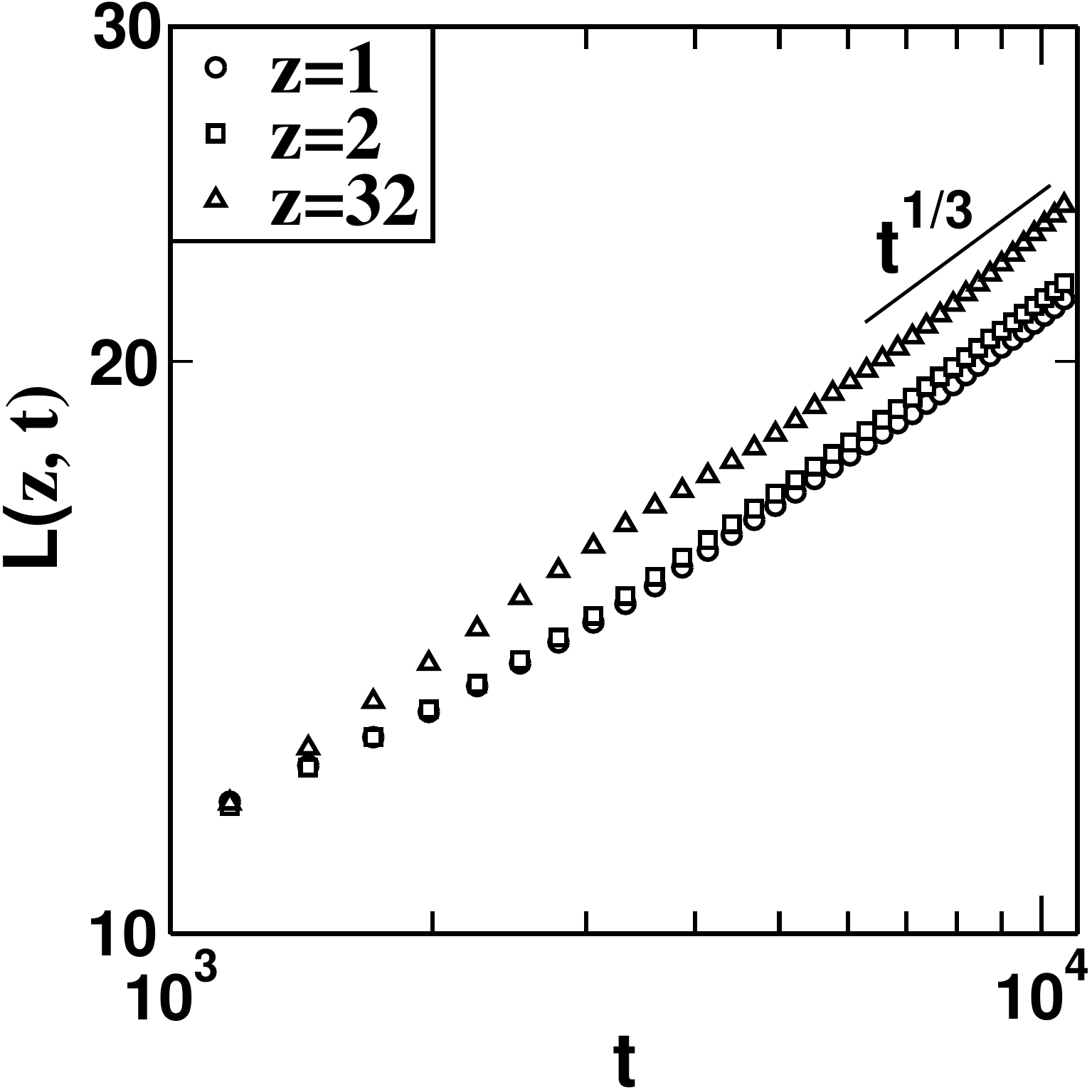}
\caption{\label{f11}SDSD on a random substrate: log-log plot of $L(z, t)$ vs. $t$ for $z=1,2,32$.}
\end{figure}

In Fig.~\ref{f11}, we show the time-dependence of the layer-wise length scales for $z=1,2,32$. In the surface 
layer, the growth is significantly slower than the bulk. However, the growth law is consistent with the LS law for all 3 layers -- only the prefactors are different. We see no signature of the logarithmic growth reported for domain growth in the {\it random-field Ising model} (RFIM) \cite{clm12,kbp17}. We expect that logarithmic growth would be observed here if the surface gives rise to a long-range random field, rather than the short-range field considered here.

\section{Summary and Discussion}
\label{sec4}

Let us conclude this paper with a summary and discussion of our results. We have studied surface-directed spinodal decomposition (SDSD) on a chemically patterned substrate. We primarily consider the case of a checkerboard pattern (see Fig.~\ref{f1}), where alternating patches are wetted by the components A and B of the mixture, respectively. However, most of our results apply to arbitrary surface patterns. In contrast to earlier studies, our goal is to make quantitative statements about the pattern dynamics near and at the surface.

We modeled the system using the Puri-Binder (PB) model of SDSD \cite{pb92}. In the PB model, bulk phase separation is described by the Cahn-Hilliard-Cook (CHC) equation with an additional term due to the surface potential. This is a fourth-order partial differential equation, so it must be supplemented by two boundary conditions whenever a surface is introduced. The first boundary condition relaxes the order parameter at the surface to its equilibrium value via nonconserved kinetics. The second boundary condition is the {\it no-flux} or {\it zero-current} condition, which accounts for the absence of material transport across the surface.

We start our simulation with a homogeneous mix of A and B, and quench the system to low temperatures. The surface patches are rapidly wetted by the preferred components on a time-scale which is much faster than that of phase separation. In the initial stages, the surface registry extends several layers into the bulk. However, the ongoing phase separation melts the surface pattern, starting with the uppermost layers. The melting process destroys the surface registry, until only the surface layer remains registered to the base pattern. The melting time ($t_c$) scales with the patch size ($M_x$) as $t_c \sim M_x^3$. Thus, the morphology at a fixed $z$ (near the substrate) shows a crossover from {\it surface registry} to {\it bulk phase separation}. This crossover can also be seen in the layer-wise correlation function $C(\rho,z,t)$ or structure factor $S(k_\rho,z,t)$, and the corresponding length scale $L(z,t)$. A universal feature that survives the crossover is the Porod tail in the structure factor, which arises due to scattering from sharp interfaces -- these are present in both the surface pattern and the bulk segregation pattern.

As stressed in the introduction, the problem of SDSD on chemically patterned substrates is of great scientific and technological importance. We hope that the theoretical results presented here will be subjected to experimental test. There are many aspects of this problem which remain poorly understood, e.g., mixture composition, hydrodynamic velocity fields in fluid mixtures, confined geometries like thin films and wedges, etc. We believe that future experiments and theoretical studies should focus on some of these outstanding problems. \\
\ \\
\noindent{\bf Acknowledgments:} PD acknowledges financial support from the Council of Scientific and Industrial Research, India.


\begin{thebibliography}{99}

\bibitem{pw09} S. Puri and V.K. Wadhawan (eds.), \textit{Kinetics of Phase Transitions}, CRC Press, Boca Raton (2009).
 
\bibitem{dp04} S. Dattagupta and S. Puri, \textit{Dissipative Phenomena in Condensed Matter: Some Applications}, Springer-Verlag, Heidelberg (2004).
 
\bibitem{ao02} A. Onuki, \textit{Phase Transition Dynamics}, Cambridge University Press, Cambridge (2002).

\bibitem{hh77} P.C. Hohenberg and B.I. Halperin, Rev. Mod. Phys. \textbf{49}, 435 (1977).
 
\bibitem{ls61} I.M. Lifshitz and V.V. Slyozov, J. Phys. Chem. Solids \textbf{19}, 35 (1961).
 
\bibitem{dh86} D.A. Huse, Phys. Rev. B \textbf{34}, 7845 (1986).
 
\bibitem{jnk91} R.A.L. Jones, L.J. Norton, E.J. Kramer, F.S. Bates and P. Wiltzius, Phys. Rev. Lett. \textbf{66}, 1326 (1991).

\bibitem{gk95} G. Krausch, Mater. Sc. Eng. R Rep. \textbf{14}, 1 (1995).
 
\bibitem{gk03} M. Geoghegan and G. Krausch, Prog. Polym. Sci. \textbf{28}, 261 (2003).
 
\bibitem{pf97} S. Puri and H.L. Frisch, J. Phys.: Condens. Matter \textbf{9}, 2109 (1997).
 
\bibitem{kb98} K. Binder, J. Non-Equilib. Thermodyn. \textbf{23}, 1 (1998).
 
\bibitem{sp05} S. Puri, J. Phys.: Condens. Matter \textbf{17}, R101 (2005).
 
\bibitem{kpd10} K. Binder, S. Puri, S. K. Das and J. Horbach, J. Stat. Phys. \textbf{138}, 51 (2010).
 
\bibitem{ty05} T. Young, Philos. Trans. R. Soc. London \textbf{95}, 65 (1805).

\bibitem{pb92} S. Puri and K. Binder, Phys. Rev. A \textbf{46}, R4487 (1992); Phys. Rev. E \textbf{49}, 5359 (1994).

\bibitem{pbzp92} S. Puri and K. Binder, Z. Phys. B \textbf{86}, 263 (1992).

\bibitem{pb94} S. Puri and K. Binder, J. Stat. Phys. \textbf{77}, 145 (1994).

\bibitem{pbf97} S. Puri, K. Binder and H.L. Frisch, Phys. Rev. E \textbf{56}, 6991 (1997).
 
\bibitem{pb01} S. Puri and K. Binder, Phys. Rev. Lett. \textbf{86}, 1797 (2001); Phys. Rev. E {\bf 66}, 061602 (2002).

\bibitem{be90} R.C. Ball and R.L.H. Essery, J. Phys. Condens. Matter \textbf{2}, 10303 (1990).

\bibitem{bc92} G. Brown and A. Chakrabarti, Phys. Rev. A \textbf{46}, 4829 (1992).

\bibitem{jm93} J.F. Marko, Phys. Rev. E \textbf{48}, 2861 (1993).

\bibitem{kbw94} A. Kumar, H.A. Biebuyck and G.M. Whitesides, Langmuir \textbf{10}, 1498 (1994).
  
\bibitem{jw02} W.C. Johnson and S. M. Wise, Appl. Phys. Lett. \textbf{81}, 919 (2002).
 
\bibitem{rjh09} A. Reina, X. Jia, J. Ho, D. Nezich, H. Son, V. Bulovic, M.S. Dresselhaus and J. Kong,  Nano Letters \textbf{9}, 30 (2009).
 
\bibitem{rxd04} E.A. Roth, T. Xu, M. Das, C. Gregory, J.J. Hickman and T. Boland, Biomaterials \textbf{25}, 3707 (2004).
 
\bibitem{cjt01} H.G. Craighead, C.D. James and A.M.P. Turner, Current Opinion in Solid State Physics and Material Science \textbf{5}, 177 (2001).
 
\bibitem{jdk98} C.D. James, R. C. Davis, L. Kam, H.G. Craighead, M. Isaacson, J.N. Turnur and W. Stain, Langmuir \textbf{14}, 741 (1998).
 
\bibitem{lwb02} Y.L. Loo, R.L. Willett, K.W. Baldwin and J.A. Rogers, Appl. Phys. Lett. \textbf{81}, 562 (2002).
 
\bibitem{jzg02} X. Jiang, H. Zheng, S. Gourdin and P.T. Hammond, Langmuir \textbf{18}, 2607 (2002).
 
\bibitem{kdl98} A. Karim, J.F. Douglas, B.P. Lee, S.C. Glotzer, J.A. Rogers, R.J. Jackman, E.J. Amis and G.M. Whitesides, Phys. Rev. E {\bf 57}, R6273 (1998).

\bibitem{end98} B.D. Ermi, G. Nisato, J.F. Douglas, J.A. Rogers and A. Karim, Phys. Rev. Lett. {\bf 81}, 3900 (1998).

\bibitem{bwm98} M. B\"oltau, S. Walheim, J. Mlynek, G. Krausch and U. Steiner, Nature \textbf{391}, 877 (1998).

\bibitem{mg99} M. Grunze, Science \textbf{283}, 41 (1999).
 
\bibitem{dbs98} E. Delamarche, A. Bernard, H. Schmid, A. Bietsch, B. Michel and H. Biebuyck, J. Am. Chem. Soc. \textbf{120}, 500 (1998).
 
\bibitem{kyb02} O. Kuksenok, J.M. Yeomans and A.C. Balazs, Phys. Rev. E \textbf{65}, 031502 (2002).
 
\bibitem{kb03} O. Kuksenok and A.C. Balazs, Phys. Rev. E \textbf{68}, 011502 (2003).
 
\bibitem{cc98} H. Chen and A. Chakrabarti, J. Chem. Phys. {\bf 108}, 6897 (1998).
 
\bibitem{os87} Y. Oono and Y. Shiwa, Mod. Phys. Lett. B \textbf{1}, 49 (1987).

\bibitem{ob88} Y. Oono and M. Bahiana, Phys. Rev. Lett. {\bf 61}, 1109 (1988).

\bibitem{wd06} X.-F. Wu and Y.A. Dzenis, J. Chem. Phys. {\bf 125}, 174707 (2006).

\bibitem{dpz13} R. Dessi, M. Pinna and A.V. Zvelindovsky, Macromolecules {\bf 46}, 1923 (2013).

\bibitem{clx13} P. Chen, H. Liang, R. Xia, J. Qian and X. Feng, Macromolecules {\bf 46}, 922 (2013).

\bibitem{spz16} M. Serral, M. Pinna, A.V. Zvelindovsky and J.B. Avalos, Macromolecules {\bf 49}, 1079 (2016).

\bibitem{hm17} J.D. Hill and P.C. Millett, Sci. Rep. {\bf 7}, 1 (2017).

\bibitem{xzw19} W. Xiang, Z. Zhu, K. Wang and L. Zhou, Phys. Chem. Chem. Phys. {\bf 21}, 641 (2019).

\bibitem{zl19} W. Zhao and W. Li, Phys. Chem. Chem. Phys. {\bf 21}, 18525 (2019).

\bibitem{po88} S. Puri and Y. Oono, J. Phys. A {\bf 21}, L755 (1988).

\bibitem{op87} Y. Oono and S. Puri, Phys. Rev. Lett. \textbf{58}, 836 (1987); Phys. Rev. A \textbf{38}, 434 (1988); S. Puri and Y. Oono, Phys. Rev. A \textbf{38}, 1542 (1988).
  
\bibitem{gp82} G. Porod, in \textit{Small-Angle X-Ray Scattering}, ed. by O. Glatter and O. Kratky, Vol. 42, Academic Press, New York (1982).

\bibitem{op88} Y. Oono and S. Puri, Mod. Phys. Lett. B \textbf{2}, 861 (1988).

\bibitem{bs84} H.D. Bale and P.W. Schmidt, Phys. Rev. Lett. {\bf 53}, 596 (1984).

\bibitem{db00} D. Das and M. Barma, Phys. Rev. Lett. {\bf 85}, 1602 (2000); M. Barma, Eur. Phys. J. B {\bf 64}, 387 (2008).

\bibitem{skb11} G.P. Shrivastav, S. Krishnamoorthy, V. Banerjee and S. Puri, Europhys. Lett. {\bf 96}, 36003 (2011).

\bibitem{skb14} G.P. Shrivastav, M. Kumar, V. Banerjee and S. Puri, Phys. Rev. E {\bf 90}, 032140 (2014).

\bibitem{clm12} F. Corberi, E. Lippiello, A. Mukherjee, S. Puri and M. Zannetti, Phys. Rev. E {\bf 85}, 021141 (2012).

\bibitem{kbp17} M. Kumar, V. Banerjee and S. Puri, Europhys. Lett. {\bf 117}, 10012 (2017).

\end{thebibliography}
\end{document}